\documentclass[manuscript]{aastex}

\usepackage{natbib}

\shorttitle{Time Dilation of SN 1997ex}
\shortauthors{Foley et al.}

\begin{document}


\title{A Definitive Measurement of Time Dilation in the Spectral
Evolution of the Moderate-Redshift Type Ia Supernova 1997ex}


\author{Ryan J. Foley\altaffilmark{1}, Alexei
V. Filippenko\altaffilmark{1}, Douglas
C. Leonard\altaffilmark{2,3}, Adam
G. Riess\altaffilmark{4}, Peter Nugent\altaffilmark{5}, and Saul
Perlmutter\altaffilmark{5}}

\altaffiltext{1}{Department of Astronomy, University of California, Berkeley,
CA 94720-3411; email rfoley@astro.berkeley.edu,
alex@astro.berkeley.edu}
\altaffiltext{2}{Astronomy Department, MS 105-24, California Institute
of Technology, Pasadena, CA 91125; email leonard@astro.caltech.edu}
\altaffiltext{3}{NSF Astronomy and Astrophysics Postdoctoral Fellow.}
\altaffiltext{4}{Space Telescope Science Institute, 3700 San Martin
Drive, Baltimore, MD 21218; email ariess@stsci.edu}
\altaffiltext{5}{Lawrence Berkeley National Laboratory, Berkeley, CA  94720;
email saul@lbl.gov, nugent@lbl.gov}



\begin{abstract}
We have obtained high-quality Keck optical spectra at three epochs of
the Type Ia supernova 1997ex, whose redshift $z$ is 0.361. The elapsed
calendar time between the first two spectra was 24.88~d, and that
between the first and third spectra was 30.95~d.  In an expanding
universe where $1 + z$ represents the factor by which space has
expanded between the emission and detection of light, the amount of
aging in the supernova rest frame should be a factor of $1 / (1 + z)$
smaller than the observed-frame aging; thus, we expect SN~1997ex to
have aged 18.28~d and 22.74~d between the first epoch and the second
and third epochs, respectively.  The quantitative method for
determining the spectral-feature age of a SN~Ia, developed by
\citet{Riess97:timedilation}, reveals that the corresponding elapsed
times in the supernova rest frame were $16.97 \pm 2.75$~d and $18.01
\pm 3.14$~d, respectively.  This result is inconsistent with no time
dilation with a significance level of 99.0\%, providing evidence
against ``tired light'' and other hypotheses in which no time dilation
is expected. Moreover, the observed timescale of spectral evolution is
inconsistent with that expected in the ``variable mass theory."  The
result is within $\sim$1$\sigma$ of the aging expected from a
universe in which redshift is produced by cosmic expansion.
\end{abstract}





\section{Introduction}\label{s:intro}

It has long been accepted that the redshifts of distant extragalactic
objects are caused predominantly by a cosmological expansion.
However, assertions that redshifts are not of cosmological origin are
occasionally made \citep[][and references therein]{Arp87,Burbidge04},
and alternative theories have been proposed that try to explain the
redshifts through other means such as ``tired light''
\citep[e.g.,][]{LaViolette86,Crawford99} and variable mass
\citep[e.g.,][]{Narlikar97}.  Despite the predominance of the current
redshift paradigm, little direct observational evidence exists in
support of cosmological expansion \citep{Sandage91, Pahre96}.


One prediction of cosmological expansion is that time will be dilated
by a factor of $1 + z$.  Most astronomical phenomena evolve on very
long time scales, making a time dilation measurement infeasible.
However, supernovae (SNe), which rise and fall in brightness on a time
scale of a few months and are visible at moderate to high redshifts,
are ideal for this experiment. Indeed, decades ago \citet{Wilson39}
and \citet{Rust74} suggested that the light curves of SNe~Ia might be
used to detect the expected time dilation.

\citet{Leibundgut96} showed that the light curve of SN~1995K ($z =
0.479$) was abnormally wide for a SN~Ia if time dilation was ignored.
However, the light curve of SN~1995K compressed by a factor of
$1/(1+z)$ provided a good match to low-$z$ SN~Ia light curves.  Since
the intrinsic light curves of SNe~Ia span a wide range (roughly
0.5--1.6 times the normal width; e.g., \citealt{Goldhaber01}), the
null hypothesis of no time dilation was not completely ruled out.
\citet{Goldhaber01} showed that a sample of SN~Ia (which included 42
high-$z$ objects) light curves is consistent ($\sim$18$\sigma$)
with time dilation.  However, this result is statistical, relying on
many objects which have already had their light-curve width modified
by an intrinsic ``stretch factor," and \citet{Narlikar97} contend that
the stretching is explained in the variable mass model.  Because
there may be evolutionary factors that change a light curve's shape, a
positive time dilation result from the examination of {\it spectral
features} provides more compelling evidence for time dilation.
Moreover, for any individual SN~Ia, the intrinsic width is unknown, so
without assuming a $1+z$ dilation, the intrinsic width and dilation
can not be separated.

In addition to the predictable nature of the SN~Ia light curve, the
SN~Ia spectrum also evolves in a very reliable way
\citep[e.g.,][]{Filippenko97}.  As shown by
\citet{Riess97:timedilation}, the method of spectral-feature aging,
which compares a single-epoch spectrum to a catalog of comparison
spectra with known ages to determine a spectral-feature age (SFA), can
determine the age of a SN~Ia using a single spectrum; no light-curve
information is necessary.  Using this method,
\citet{Riess97:timedilation} demonstrated that the multi-epoch spectra
of SN~1996bj are consistent with a $(1 + z)$-stretched temporal
evolution, and inconsistent with no time dilation at the 96.4\%
confidence level.

Following the method of \citet{Riess97:timedilation}, we examine the
spectral evolution of SN~1997ex, a normal SN~Ia with $z = 0.361$, to
determine the SFA at the time the spectra were obtained.  Comparing
this result to the time between epochs will result in a measurement of
the time dilation.  We present the observations in
Section~\ref{s:observations}, and in Section~\ref{s:dilation} we
determine the rate at which SN~1997ex ages.
Section~\ref{s:discussion} discusses the results and implications of
our measurement.

\section{Summary of Observations}\label{s:observations}

SN~1997ex was discovered on 28 December 1997 (UT dates are used
throughout this paper) by the Supernova Cosmology Project
\citep{Nugent98}.  Spectra were obtained on 1998 January 1, 1998
January 26, 1998 February 1, and 1998 March 5 using the Keck~II 10-m
telescope with the Low Resolution Imaging Spectrometer
\citep[LRIS;][]{Oke95}.  We were able to determine accurate SFAs for
the first three epochs, but not for the fourth epoch.  

The original spectra were contaminated by host-galaxy light.  Removal
of this light is important for creating accurate SFAs.  We obtained a
Keck/LRIS spectrum of the host galaxy on 1999 January 21, approximately 
a year after discovery of SN~1997ex, 
with the intention of using these data for galaxy light
subtraction.  During this time the SN had faded significantly; the SN
component of this latest spectrum is below the noise level.  First, we
extracted the SN without any galaxy subtraction.  Then, scaling to
narrow lines present in both the SN spectra and the galaxy template
spectrum, we were able to subtract the galaxy component of the
original SN spectra.

\begin{figure}
\begin{center}
\rotatebox{90}{
\scalebox{0.6}{
\plotone{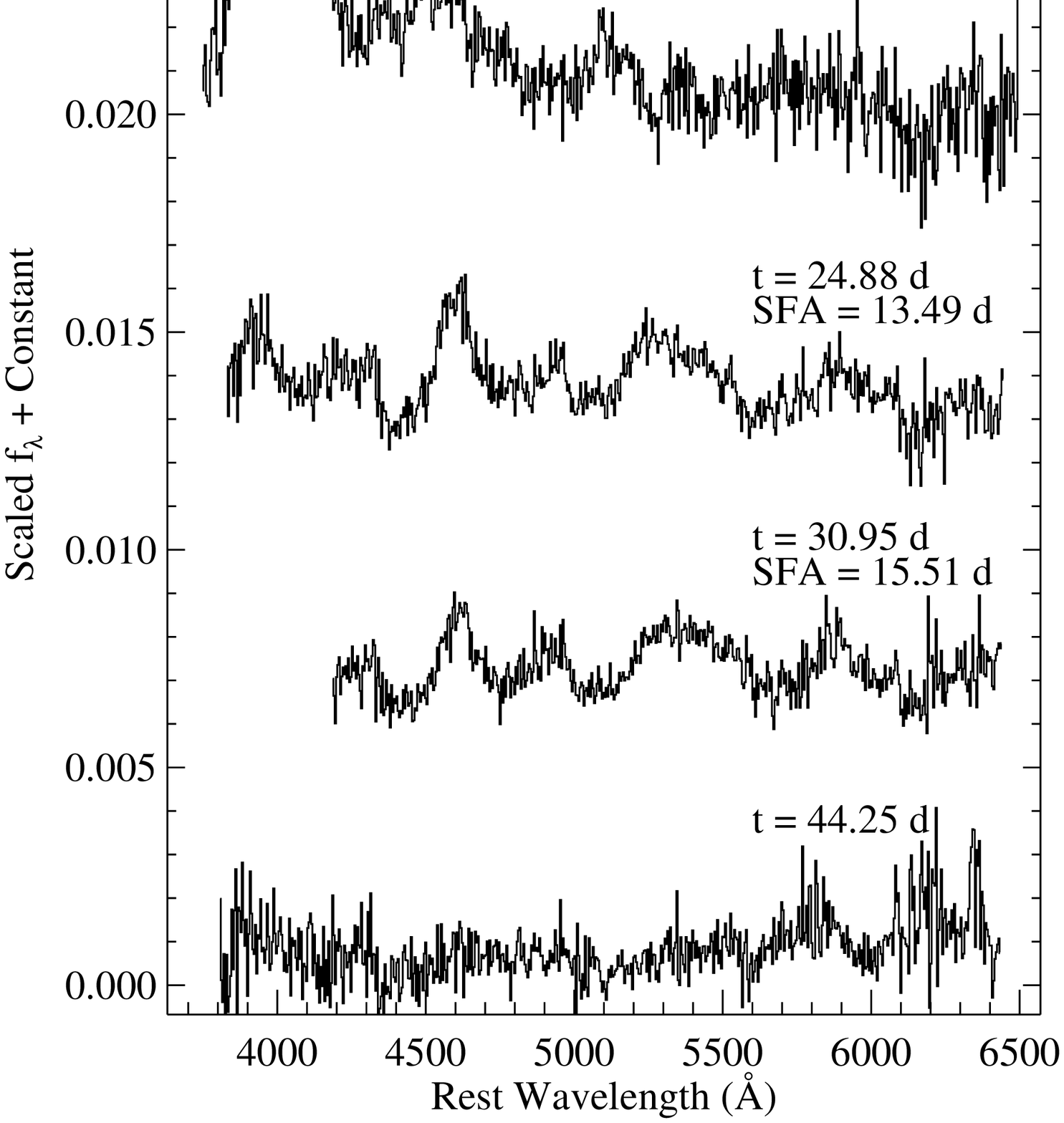}
}
}
\end{center}
\caption{Spectra of SN~1997ex.  The spectra have been deredshifted to
the rest frame, and host-galaxy light has been subtracted.}\label{f:allspec}
\end{figure}

\begin{figure}
\begin{center}
\rotatebox{90}{
\scalebox{0.6}{
\plotone{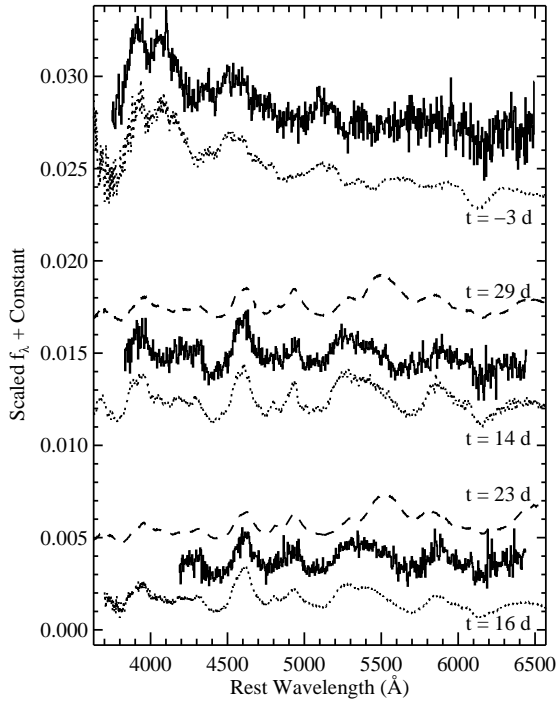}
}
}
\end{center}
\caption{Spectra of SN~1997ex with low-$z$ SN comparison spectra.  The
spectra have been deredshifted to the rest frame.  The dotted lines
are spectra at the SFA for each SN~1997ex spectrum (SN~1995E
\citep{Riess98:lambda}, SN~1990N (Filippenko, private communication),
and SN~1994D \citep{Filippenko97} are at photometric ages of $-3$~d,
14~d, and 16~d relative to $B$~maximum, respectively).  The dashed
lines are spectra of SN~1994D at the epoch corresponding to the time
elapsed in the observed frame, assuming that the first-epoch SFA is
equivalent to its true age ($t = -2.28$~d).  Clearly, the latter
spectra are a poor match to the observed spectra of
SN~1997ex.}\label{f:speccomp}
\end{figure}

The deredshifted, galaxy-subtracted spectra are shown in
Figure~\ref{f:allspec}.  Figure~\ref{f:speccomp} shows the rest-frame
spectra of the three epochs of SN~1997ex along with low-$z$ comparison
spectra.  All comparison spectra are of ``Branch-normal SNe~Ia''
\citep{Branch93:normal}.  Examining the earliest spectrum, we see
strong absorption on the red side of the \ion{Ca}{2} H \& K lines and
the lack of \ion{Ti}{2} absorption at $\sim$4200~\AA, indicating
that SN~1997ex is not similar to the subluminous SN~1991bg
\citep[e.g.,][]{Filippenko92:91bg} or the overluminous SN~1991T
\citep[e.g.,][]{Filippenko92:91T}.  Although pre-maximum spectra are
particularly useful for identifying the peculiar nature of a SN~Ia, we
note that the late-time spectra of SN~1997ex are consistent with a
Branch-normal SN~Ia as well.

\section{Time Dilation Measurement from Spectral-Feature
Aging}\label{s:dilation}

The observations of SN~1997ex were separated by 24.88~d (between the
first and second epochs), 6.07~d (between the second and third
epochs), and 31.92~d (between the third and fourth epochs).  In an
expanding universe, the spectra should have aged only 18.28~d, 4.46~d,
and 23.45~d (respectively) between these epochs.  Using the
spectral-feature aging method of \citet{Riess97:timedilation}, we have
determined the SFA for three of our four spectra, as presented
in Table~\ref{t:ages}.  For this calculation, we have used the
features listed in Table~\ref{t:features}.  These features were
selected by a genetic algorithm designed to choose the best subset of
features from a randomly generated set of 400 features.  This method
chooses different subsets of features, keeps the subsets with the best
age determinations, redistributes the remaining features into new
subsets and iterates. (see \citealt{Charbonneau95} for a general
description and \citealt{Foley05} for details).

The fourth spectrum has a lower signal-to-noise ratio (S/N) than the
other spectra.  Moreover, the age of the SN at the fourth epoch is
larger than the age of any SN spectrum in the catalog invoked for the
spectral-feature aging method \citep{Riess97:timedilation}.  The
spectrum was obtained 62.87~d after the first-epoch spectrum and
$59.30 \pm 1.66$~d after the time of $B$~maximum found from fitting
the SFAs (the date corresponding to $t = 0$ in the fit). Even with the
time dilation corresponding to $z = 0.361$, the fourth-epoch spectrum
should have an SFA of $43.57 \pm 1.22$~d.  Since the training set of
\citet{Riess97:timedilation} has one spectrum at $t = 40$~d and none
of older spectra and the S/N of the fourth-epoch spectrum is low, it
is understandable that no reliable age for this spectrum was obtained.

Between the first two epochs (an observed time of 24.88~d), the SFA
changed by $16.96 \pm 2.75$~d.  Between the second and third epochs
(an observed time of 6.07~d), the SFA changed by only $1.04 \pm
3.03$~d.

Shifting the data in Table~\ref{t:ages} so that the SFAs have a
weighted mean of zero, we were able to use a least-squares method to
fit the data with no covariance between the age factor and the date of
$B$ maximum, resulting in an age factor of $1.602 \pm 0.234$.  Since
the no time dilation model predicts an age factor of 1.000, this age
factor corresponds to a $2.57 \sigma$ result if there were no time
dilation; thus, the measured aging for SN~1997ex is inconsistent with
the null hypothesis at the 99.0\% confidence level.

The observed age factor is larger than the expected factor of $1 + z
= 1.361$.  The difference between the observed age factor and the
expected value corresponds to a $1.03 \sigma$ event, which should be
observed 30.3\% of the time by chance.

\section{Discussion}\label{s:discussion}

We have shown that the spectral evolution of SN~1997ex has likely been
dilated by a factor of $1 + z$, as expected in an expanding
universe.  This result also shows that the null hypothesis of no
cosmic expansion is ruled out at the 97.9\% level.  Thus, ``tired
light'' and other hypotheses that predict no time dilation are
essentially eliminated.  To be consistent with the observations,
alternatives such as the ``variable mass theory"
\citep[e.g.,][]{Narlikar97} would require a highly unlikely series of
coincidences, culminating with SN~Ia spectral evolution that mimics
the result expected with simple time dilation.  \citet{Narlikar97}
discount the results of \citet{Leibundgut96} since the variable mass
theory changes the decay rate of $^{56}$Ni (which dictates the shape
of a SN~Ia light curve) with redshift.  However, the evolution of
spectral features depends on the composition of the ejecta and
opacities, and therefore on temperature and density.  Although the
redshift of a given spectral feature will depend on the mass of
subatomic particles, the evolution of that line does not depend on the
mass of these particles.  Since spectral evolution should not change
with redshift, we therefore can rule out the variable mass
theory with our current data.

With the current large-scale searches for high-redshift SNe, such as
ESSENCE\\ (http://www.ctio.noao.edu/wproject;\\ \citealt{Smith02}) and the
CFHT Legacy Survey (http://www.cfht.hawaii.edu/Science/CFHLS;
\citealt{Pain02:CFHTLS}), there are many more multi-epoch
high-redshift SN~Ia spectra.  A similar analysis should be performed
on this larger data set to confirm and improve upon these results.
Knowing the time of rest-frame $B$ maximum is helpful in two ways: (1)
this time is independent of any intrinsic width of the SN light curve,
avoiding many of the photometric issues found in \citet{Leibundgut96};
and (2) this time is well-constrained and avoids any evolutionary
effects of SNe~Ia.  The time of $B$ maximum is another point that can
be used to determine time dilation.

Given a large sample of SNe~Ia, well-sampled light curves, and at least
one spectrum of each SN, the effect of time dilation across this sample
could be easily determined.  One can place all of the SFAs and
spectral epochs on the same axes by matching the time of $B$ maximum
and scaling by redshift.  With a single multi-epoch SN, we were able
to exclude the null hypothesis at the 97.9\% level; with a hundred
data points, one should be able to easily rule out the null hypothesis
at a much higher significance level.

An improvement in the SN~Ia time dilation measurement can be obtained
through higher S/N spectra (reducing the errors on an individual SFA),
more epochs per SN, an expanded training set for spectral-feature
aging (to reduce the inherent SFA errors and to match earlier-time and
later-time spectra), higher-redshift SNe (which should show even more
time dilation), and more objects.  The increase in the number of
objects is certain to occur as the major high-redshift SN~Ia searches
are finding $\sim$100 SNe~Ia per year.  The higher S/N spectra
and multiple-epoch data sets are difficult to obtain because of the
constraints on spectroscopic follow-up time of SNe~Ia on
large-aperture telescopes.  Likewise, much time must be spent to
obtain even a single spectrum of a faint SN~Ia at $z \approx 1$.
However, a better training set will be available soon, after more light
curves are published (CfA sample, \citealt{Jha05}; LOSS sample,
\citealt{Li05}).

\begin{acknowledgments}
We thank the Keck Observatory staff for their assistance, and S.
Deustua, I. Hook, R. Knop, and E. Moran for their help with some of
the observations. The W.~M. Keck Observatory is operated as a
scientific partnership among the California Institute of Technology,
the University of California, and NASA; the observatory was made
possible by the generous financial support of the W.~M. Keck
Foundation.  A.V.F. is grateful for the support of NSF grants
AST-0307894 and AST-0443378, and for a Miller Research Professorship
at UC Berkeley during which part of this work was completed.
D.C.L. is supported by a National Science Foundation (NSF) Astronomy
and Astrophysics Postdoctoral Fellowship under award AST-0401479.
\end{acknowledgments}

\bibliographystyle{apj}
\bibliography{astro_refs}


\clearpage

\clearpage

\clearpage

\begin{figure}
\begin{center}
\rotatebox{90}{
\scalebox{0.5}{
\plotone{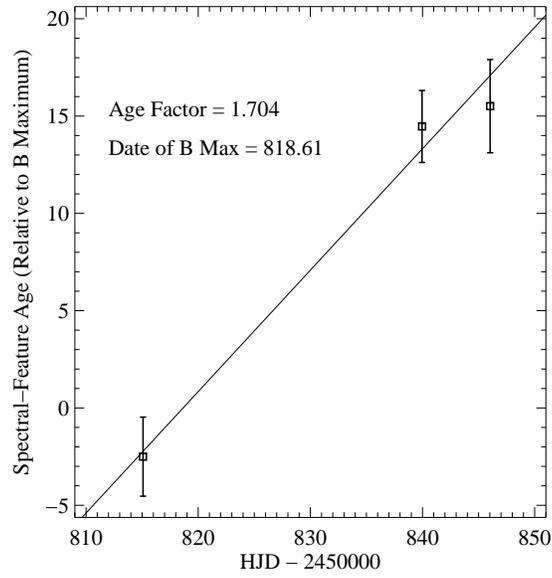}
}
}
\end{center}
\caption{Age-factor fit to spectral-feature ages.  The slope of the
fit corresponds to an age factor of $1.704 \pm 0.320$.  $B$-band maximum
is fit to be HJD $2450818.61 \pm 2.09$.}\label{f:fit}
\end{figure}


\clearpage

\begin{deluxetable}{lcr}
\tabletypesize{\scriptsize}
\tablewidth{0pt}
\tablecaption{Spectral Feature Ages\label{t:ages}}
\tablehead{
\colhead{UT Date} &
\colhead{HJD- 2450000} &
\colhead{SFA (days)} }

\startdata

1998-01-01 & 815.08 & $-2.50$ (2.03)\\

1998-01-26 & 839.96 & 14.46 (1.85)\\

1998-02-01 & 846.03 & 15.51 (2.39)$\!$



\tablecomments{Spectral-feature age (SFA) is relative to $B$-band
maximum.  Uncertainties are indicated in parentheses.}

\enddata

\end{deluxetable}

\clearpage

\begin{deluxetable}{cc}
\tabletypesize{\scriptsize}
\tablewidth{0pt}
\tablecaption{Spectral Features\label{t:features}}
\tablehead{
\colhead{Feature} &
\colhead{Wavelength Range (\AA)} }

\startdata

1 & 4397.72--5871.53 \\
2 & 4682.52--5122.65 \\
3 & 4725.47--4999.27 \\
4 & 5593.93--5910.65 \\
5 & 5642.64--6249.96 \\
6 & 5716.59--6193.18 \\
7 & 5719.52--5845.31 \\
8 & 6015.48--6131.82

\enddata

\end{deluxetable}

\clearpage

\begin{deluxetable}{lcccccccccccl}
\tabletypesize{\scriptsize}
\tablewidth{0pt}
\tablecaption{Journal of Spectroscopic Observations of SN 1997ex\label{t:dates}}
\tablehead{\colhead{UT Date} &
\colhead{HJD} &
\colhead{Tel.\tablenotemark{a}} &
\colhead{Range\tablenotemark{b}}  &
\colhead{Res.\tablenotemark{c}} &
\colhead{P.A.\tablenotemark{d}} &
\colhead{O.P.A.\tablenotemark{e}} &
\colhead{Air.\tablenotemark{f}} &
\colhead{F. Std.\tablenotemark{g}} &
\colhead{See.\tablenotemark{h}} &
\colhead{Slit} &
\colhead{Exp.} &
\colhead{Observers\tablenotemark{i}} \\
\colhead{yymmdd} &
\colhead{$-$2450000} &
\colhead{} &
\colhead{(\AA)} &
\colhead{(\AA)} &
\colhead{(deg)} &
\colhead{(deg)} &
\colhead{} &
\colhead{} &
\colhead{($''$)} &
\colhead{($''$)} &
\colhead{(s)} &
\colhead{} }

\startdata

980101 & 815.08 & KII & 4900-9860 & 10 & 124 & 58 & 1.17 & HD84 &
1.0 & 1 & 600 & P,H,K,D \\

980126 & 839.96 & KII & 5184-8958 & 7 & 40 & 34 & 1.05 & BD26 & 1.1
& 1 & 3x1200 & F,L,R \\

980201 & 846.03 & KII & 5696-9486 & 7 & 60 & 66 & 1.36 & HD19 & 1.5
& 1 & 2x1200,1300 & F,L,R \\

980305 & 877.95 & KII & 5180-8946 & 7 & 63 & 66 & 1.36 & BD26 & 1.1
& 1 & 2x1800 & F,M \\

990121 & 1199.90 & KII & 5100-8840 & 7 & 63 & 129 & 1.10 & HD19 &
0.8 & 1 & 2x1200 & F,L \\

\enddata

\tablecomments{Additional observation details: 980101: 300/5000
grating.  Extracted $\pm 5$ pixels, background $\pm 7$--20.  980126:
400/8500 grating.  Extracted $\pm 5$ pixels, background $\pm 8$--18.
Dithered between exposures.  980201: 400/8500 grating.  Extracted $\pm
5$ pixels, background $\pm 10$--25. 980305: 400/8500 grating.
Extracted $\pm 5$ pixels, background $\pm 10$--25.  990121: Extracted
$\pm 5$ pixels, background $\pm 10$--25.  In all cases the CCD pixel
scale was $0.42''$~pixel$^{-1}$, and an optimal extraction
\citep{Horne86} was performed.}

\tablenotetext{a}{KII = Keck-II 10-m/Low Resolution Imaging
Spectrometer \citep{Oke95}.}

\tablenotetext{b}{Observed wavelength range of spectrum.  In some
cases, the ends are very noisy, and are not shown in the figures.}

\tablenotetext{c}{Approximate spectral resolution derived from
night-sky lines.}

\tablenotetext{d}{Position angle of the spectrograph slit.  }

\tablenotetext{e}{Optimal parallactic angle \citep{Filippenko82} near
the midpoint of the exposures.}

\tablenotetext{f}{Average airmass of observations.}

\tablenotetext{g}{The standard stars are as follows: HD19 = HD~19445, HD84 =
HD~84937, BD26 = BD+26$^\circ$2606 \citep{Oke83}.}

\tablenotetext{h}{Seeing is estimated from the FWHM of point sources on the CCD
chip.}

\tablenotetext{i}{D = Susana Deustua; F = Alex Filippenko; H =
Isobel Hook; K = Robert Knop; L = Douglas Leonard; M = Ed Moran; P =
Saul Perlmutter; R = Adam Riess.}

\end{deluxetable}

\end{document}